\documentclass{article}
\usepackage[utf8]{inputenc}
\usepackage{amsmath,amsthm,amsfonts,amssymb}
\usepackage{fullpage}
\usepackage{multirow}
\usepackage[noblocks]{authblk}
\usepackage{chngcntr}
\usepackage{tikz}
\usepackage{float}
\usepackage[rightcaption]{sidecap}
\usepackage{enumerate}\usepackage{enumitem}
\usepackage{hanging}
\usepackage{subcaption} 
\usepackage{diagbox}
\usepackage{soul}
\usepackage{hyperref}
\usepackage{fancyvrb}
\usepackage[labelfont=bf]{caption}
\usepackage[font={small}]{caption}
\usepackage[export]{adjustbox}
\usepackage{graphicx,wrapfig,lipsum}
\usepackage{array}
\usepackage{bbm}
\usepackage{verbatim}
\usepackage{hyperref}
\usepackage[absolute]{textpos}

\newenvironment{blockquote}{
  \par
  \medskip
  \leftskip=2em\rightskip=0em
  \noindent\ignorespaces}{
  \par\medskip}
  
\hypersetup{
    colorlinks=true,
    linkcolor=blue,
    filecolor=magenta,      
    urlcolor=black,
    pdftitle={Overleaf Example},
    pdfpagemode=FullScreen,
    }

\title{Histropy: A Computer Program for Quantifications of Histograms of 2D Gray-scale Images}
\author[2]{Sagarika Menon$^*$, Peter Moeck$^{**}$}
\date{
    Nano-Crystallography Group, Department of Physics\\
    Portland State University, Portland, OR 97207-0751, $^*$smenon92@gatech.edu
    $^{**}$pmoeck@pdx.edu\\}

\begin{document}

\maketitle
\begin{abstract}
The computer program “Histropy” is an interactive Python program for the quantification of selected features of two-dimensional (2D) images/patterns (in either JPG/JPEG, PNG, GIF, BMP, or baseline TIF/TIFF formats) by means of calculations based on the pixel intensities in this data, their histograms, and user-selected sections of those histograms. The histograms of these images display pixel-intensity values along the x-axis (of a 2D Cartesian plot), with the frequency of each intensity value within the image represented along the y-axis. The images need to be of 8-bit or 16-bit information depth and can be of arbitrary size. (Up to 1024 pixels maximum on both sides is recommended as larger images tend to significantly slow the program’s performance). Histropy generates an image’s histogram surrounded by a graphical user interface that allows one to select any range of image-pixel intensity levels, i.e. sections along the histograms’ x-axis, using either the computer mouse or numerical text entries. The program subsequently calculates the (so-called Monkey Model) Shannon entropy and root-mean-square contrast for the selected section and displays them as part of what we call a “histogram-workspace-plot.” To support the visual identification of small peaks in the histograms, the user can switch between a linear and log-base-10 display scale for the y-axis of the histograms. Pixel intensity data from different images can be overlaid onto the same histogram-workspace-plot for visual comparisons. The visual outputs of the program can be saved as histogram-workspace-plots in the PNG format for future usage. The source code of the program and a brief user manual are published in the supporting materials as well as on GitHub (\url{https://github.com/SMenon-14/Histropy}). Instead of taking only 2D images as inputs, the program’s functionality could be extended by a few lines of code to other potential uses employing data tables with one or two dimensions in the CSV format.
\end{abstract}
\section{Introduction}
The term histogram was coined around 1892 by statistician Karl Pearson to describe the visual representation of the distribution of data that quantified the frequency of data that fell into “bins” of a certain range. Histograms themselves were “first conceived as a visual aid to statistical approximations” (Ioannidis, 2003). The visual analysis that a histogram facilitates, combined with the quantitative information that can be extracted from it, gives histograms a wide range of applications, ranging from analyzing the distributions of test scores in a classroom to probabilistically characterizing the behavior of river discharge. 
\\Another typical application of histograms is the analysis of digital images, used in fields ranging from microscopy (Wood, 2026) to computer vision.  Pixel intensity histograms, which represent the distribution of the intensities of all pixels of an image, provide a measure of the image that can be useful for identifying similar images, compressing the image, and more. The here briefly described computer program “Histropy'' generates a pixel intensity histogram for images that are either in grayscale, Red-Green-Blue color (RGB), or a uniform hue with a color-tone-range and allows for the numerical analysis of user-selected bin-ranges within the histogram. (One of its name-giving features is the calculation of the so-called Monkey-Model (Razlighi \& Kehtarnavaz, 2009) Shannon entropy of 2D images.)
\\The computer program’s prospective use within our research group is to quantitatively distinguish between symmetries and pseudosymmetries in a series of noisy as well as noise-filtered 2D-periodic images, i.e. crystal patterns. That noise-filtering of experimentally obtained and synthetic crystal patterns will involve crystallographic image processing (Hovmöller, Oleynikov \& Zou, 2011) after objective, i.e. information-theory-based (Kanatani, 1998), crystallographic symmetry classifications (Moeck, 2022). See appendix for an example of a highly pseudosymmetric crystal pattern. As part of such quantitative distinctions between symmetries and pseudosymmetries, both the so-called “Monkey-Model” version (Razlighi \& Kehtarnavaz, 2009) of the Shannon entropy and the standard root-mean-square (RMS) contrast in original or converted gray-scale crystal patterns need to be calculated. 
\\For gray-scale patterns such as the one shown in the background of Fig. A1a, where there are a few pronounced peaks in the corresponding histogram (Fig. A1b), it will be informative to make quantifying calculations for selected ranges of the pixel intensity values. (This may be considered as constituting a very basic form of pattern segmentation.) Competing computer programs, e.g. the histogram routines that are part of the well-known electron crystallography software CRISP (Hovmöller, Oleynikov \& Zou, 2011), do not typically offer the functionality desired for our studies. (Figure A-2 illustrates the effect of crystallographic image processing in two non-disjoint plane symmetry groups on a noisy version of the crystal pattern that is shown in the background of Fig. A1a.)
\\Because the source code of our program is freely available on GitHub [2024], other researchers are invited to download and modify it to support their own studies that do not need to be based on information in two-dimensional crystal patterns or gray-scale images. (Other applications of histograms of 2D images may arise over time as well when parts of our code get reused.) 

\section{Description of the Computer Program}
As implied by the title of the paper, gray-scale images are the standard input option. Note that Histropy can only handle images of 16-bit depth when they are gray-scale so 16-bit color images must be converted before opening in Histropy. This conversion can be done using the open access program GIMP (\url{https://www.gimp.org/}) by going to the "Image" dropdown menu and then selecting "Grayscale" for the "Mode". When the user selects an 8-bit color image to view/analyze through Histropy, the program converts the image to grayscale using the following equation to compute the intensity value for each individual pixel based on the pixel’s RGB value:
\begin{equation}
\text{Grayscale value} = 0.299 * R + 0.587 * G + 0.114 * B
\end{equation}
where R stands for red, G for green, and B for blue. The coefficients reflect the fact that humans are most sensitive to green light and least sensitive to blue.
Histropy’s histogram-workspace plot consists of the histogram of the individual pixel intensities of a user-selected 2D image (or crystal pattern), a toolbar at the bottom of the screen that allows the user to navigate the histogram, four selection spaces to the right of the histogram that allow the user to quantify the histogram and image’s information, and a display of the selected image being plotted to the right of the selection spaces, Fig. 1. Note that the Histropy window must be viewed in full-screen mode for the layout seen on the next page.
\\
When processing 8-bit images, Histropy runs all actions in one to five seconds. Processing 16-bit images is about fifteen times slower on average with most operations taking around thirty seconds. These processing times will increase as the user overlays more images. These timings were taken by running Histropy on a MacBook Pro with an Apple M3 processor, which has an 4.05 GHz max CPU clock rate.
 \begin{figure}[H]
    \label{fig:initial-setup}
    \includegraphics[width=\textwidth]{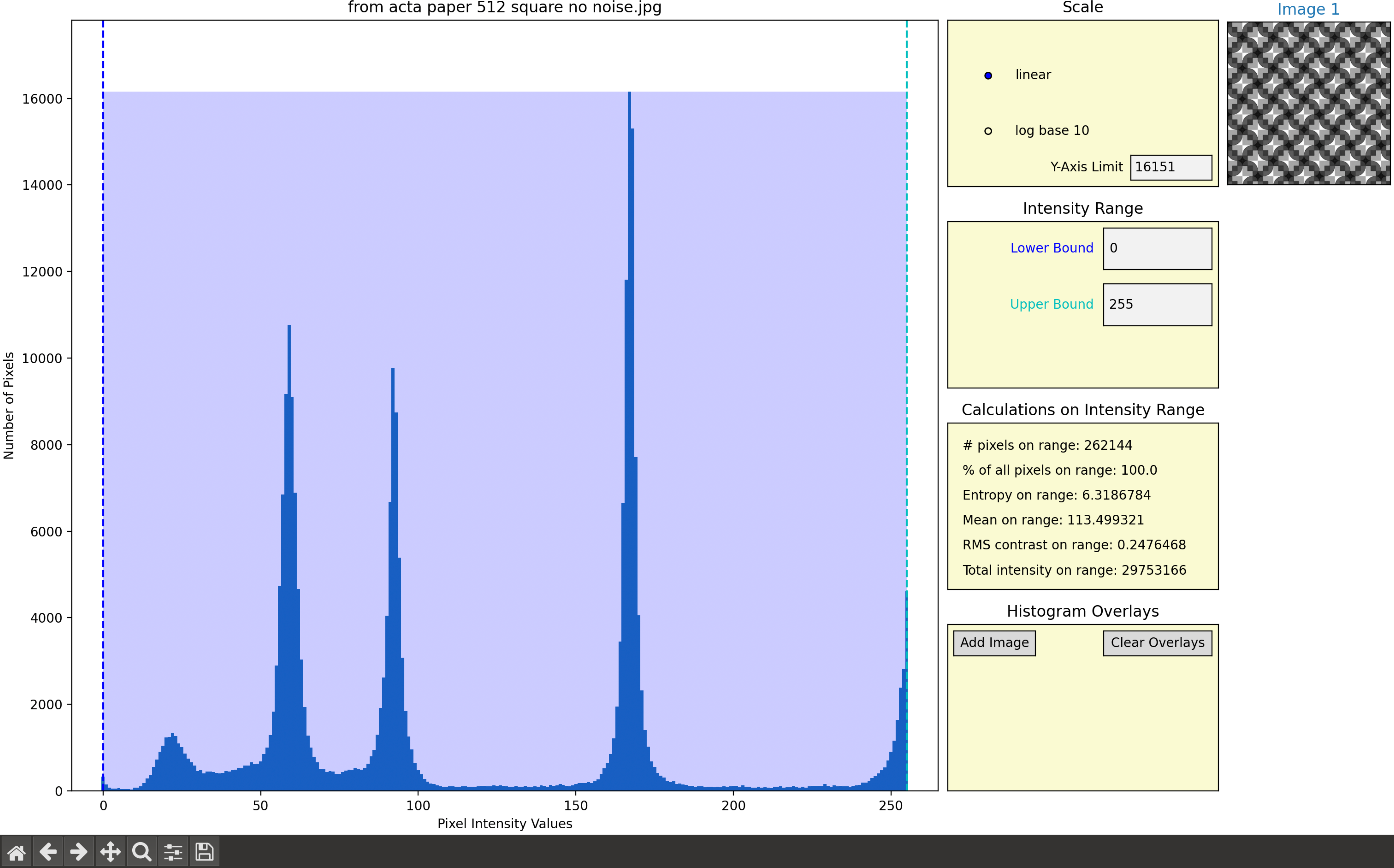}
	\caption{A screenshot of the (initial) Histropy workspace after the user has selected the image displayed in the top right corner. Note that the histogram's title contains the file name of the selected image. Also note that the title of the image in the top right corner, "Image 1," has a font color corresponding to the color of the histogram. This particular histogram is from a 512 by 512 pixel cutout of the crystal pattern in the background of Figure A1a. (This figure is displayed in color in the online version of this paper.)}
    \end{figure}
\begin{minipage}{11 cm}
\subsection{Selection Space 1: Scale}
The first selection space, “Scale,” shown in Fig. 2, allows the user to click-change between a linear and log base 10 scale, which affects the y-axis (display of number of pixels in a standard bin of unity) scale on the histogram. It also contains an input field for a y-axis limit, which defaults to the maximum y-value in the histogram. This value represents how many pixels have the most common pixel intensity value or the height of the most prominent peak in the histogram.
\subsection{Selection Space 2: Intensity Range}
The text fields in the second selection space, “Intensity Range,” displayed in Fig. 3, set the range for calculations performed in the third selection space. These text fields default to the minimum and maximum pixel intensity in the image. They can be set by directly typing into the fields themselves or by clicking directly on the histogram, which will automatically set the range in the selection space text fields to the x-value corresponding to where the user’s mouse is pointing. 
\end{minipage}
\vspace*{-7.8cm}
\begin{figure}[H]
    \includegraphics[width=4cm, right]{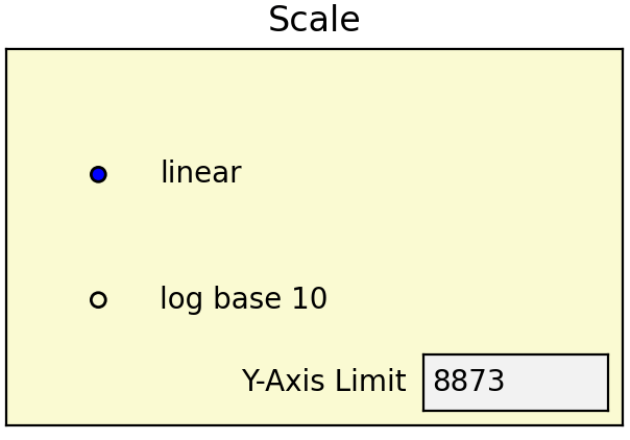} 
    \captionsetup{justification=raggedleft}
        \caption{A close up of the\\scale selection space. The \\linear y-axis is currently\\selected and the y-axis limit\\is set to the default: the max\\value of the inputted image.}
    \label{fig:scale-selection}
\end{figure}
\newpage
\begin{minipage}{11 cm}
When the user's mouse is hovering over the histogram, the text in the bottom right corner of the screen will display the coordinates that the mouse is over, Fig. 4. These coordinates can be used to accurately select the x-values for the range. The range selected is visually represented by the vertical bars and a translucent blue rectangle shown on the histogram, Fig. 1. This range selection allows the user to perform a simple form of segmentation by separating out the pixel bins that contribute to a certain histogram peak.
\end{minipage}
\vspace*{-4.1 cm}
\begin{figure}[H]
    \includegraphics[width=4cm, right]{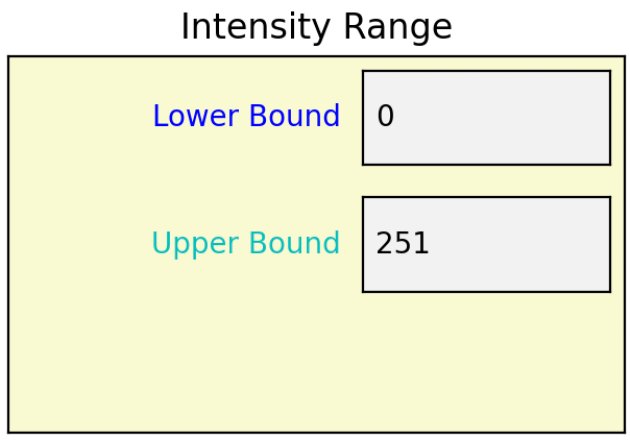} 
    \captionsetup{justification=raggedleft}
        \caption{The selection space\\for the range of pixel\\intensities (x-values on the\\histogram) that the calculations\\are computed over. The text\\fields shown here are editable\\directly by clicking on them\\and typing in new values.}
    \label{fig:range-selection-space}
\end{figure}
\begin{figure}[H]
    \label{fig:range-selection}
    \includegraphics[width=\textwidth]{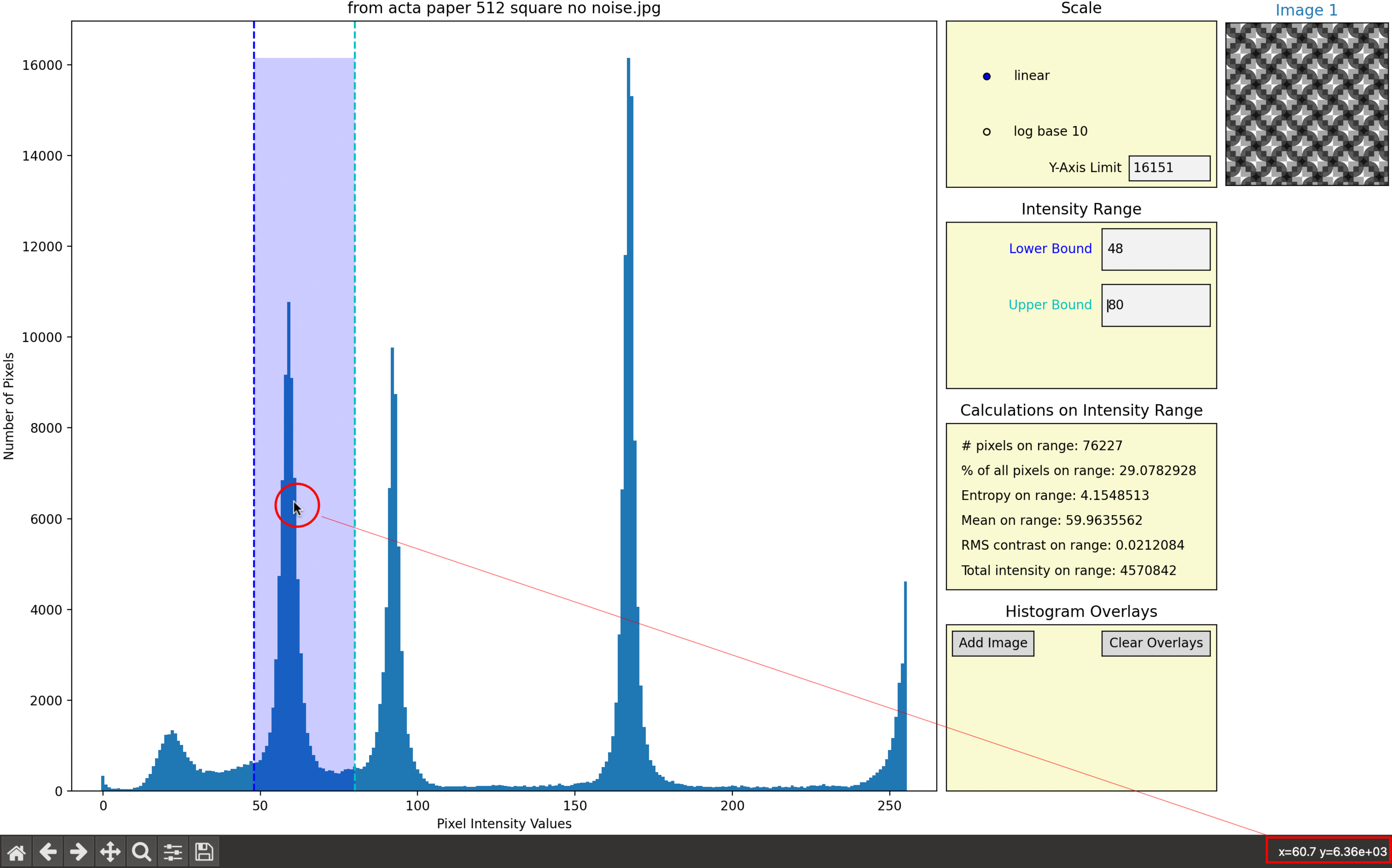}
	\caption{The cursor, circled in red on the histogram here, corresponds to the coordinates shown in the bottom right corner in a red box. The added red boxes and arrows indicate the connection between the cursor placement and the value displayed in the bottom right corner. Upon clicking at this coordinate, the lower bound of the intensity range would update to 61 (where the cursor is pointing), and this change would reflect in the selection space text box. (This figure is displayed in color in the online version of this paper.)}
\end{figure}
\newpage
\begin{minipage}{11 cm}
    \subsection{Selection Space 3: Calculations}
    The third selection space, “Calculations,” shown in Fig. 5, is a display of the following calculations over a user-selected range (which defaults to the entire range of the histogram): 
    \begin{enumerate}
        \item The number of pixels on the given range
        \item The percent of total pixels that are found in the given range
        \item A form of the Shannon entropy on the given range
        \item The mean on the given range
        \item The RMS contrast on the range
        \item The total pixel intensity on the range (a sum)
    \end{enumerate}
\end{minipage}
\vspace*{-7cm}
\begin{figure}[H]
    \includegraphics[width=4cm, right]{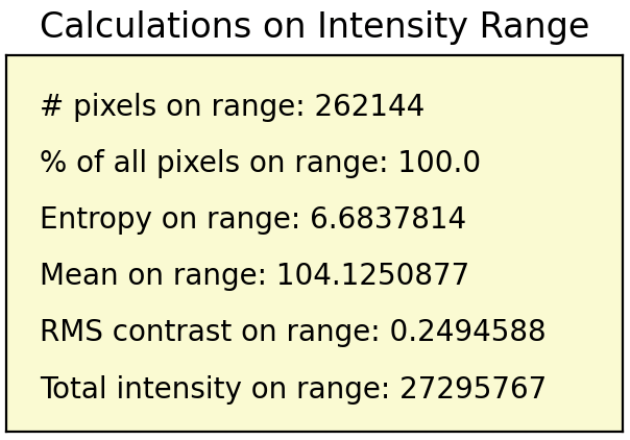} 
    \captionsetup{justification=raggedleft}
        \caption{The calculation\\selection space. These text\\ displays are not editable\\but they will update as the\\ intensity range field is\\updated, either through\\directly entering into the\\fields or by clicking on\\the histogram.}
    \label{fig:calculations-selection}
\end{figure}
\begin{blockquote}
Note that these calculations are not impacted by y-axis limitings done in the “Scale” selection space.
\\\\In the equations below, $N$ is the number of pixels on the range, $p_i$ is the intensity of the $i^{th}$ pixel, and d is $2^{\text{bit depth}} - 1$ (e.g. $d=2^8-1$ for an 8-bit image). For a 2D image’s Monkey Model Shannon entropy, $b_j$ is the number of pixels in the $j^{th}$ bin (the number of pixels with intensity/gray scale value $j$). This tells us that the probability of a pixel having value $j$ is $P(x_j)=\frac{b_j}{N}$, and therefore the one-dimensional Shannon entropy is calculated using the following equation:
\begin{equation}
    -\sum_{j=0}^{d} \frac{b_j}{N}\cdot\log_2(\frac{b_j}{N}).
\end{equation}
 The mean of the range is the sum of the pixel intensities, $p_i$, over a range divided by the number of pixels in that range:
\begin{equation}
    \bar{p} = \left(\frac{1}{N}\right)\sum_{i=1}^{N} p_i.
\end{equation}
A given range’s total image intensity is calculated as $N\cdot\bar{p}$ and its RMS contrast is calculated using
\begin{equation}
    \frac{1}{d}\sqrt{\sum_{i=1}^{N}\frac{\left(p_i-\bar{p}\right)^2}{N}}.
\end{equation}
The factor $\frac{1}{d}$ normalizes the calculation’s output to values between 0 and 1. 
\end{blockquote}
\begin{minipage}{11 cm}
    \subsection{Selection Space 4: Histogram Overlay}
    The final selection space, “Histogram Overlays,” Fig. 6, allows the user to add images whose pixel intensity data will be overlaid onto the histogram of the first image. The second image that a user adds will have its quantifying calculation data displayed in the color corresponding to its plot in the “Histogram Overlays” selection space, Fig. 7. Currently, Histropy is only capable of displaying these quantifications for one image in addition to the user’s first image. As another image is read into the program using the fourth selection space, the selection space will switch to displaying the file names of the added images in colors corresponding to how they appear on the histogram, Fig. 7. The first four images that the user overlays will appear on the right underneath the original image with their title colors corresponding to their appearance on the histogram, Fig. 8. Up to 22 images can be overlaid but it is recommended to stick to 10 overlaid images or less in order to maximize
\end{minipage}
\vspace*{-8cm}
\begin{figure}[H]
    \includegraphics[width=4cm, right]{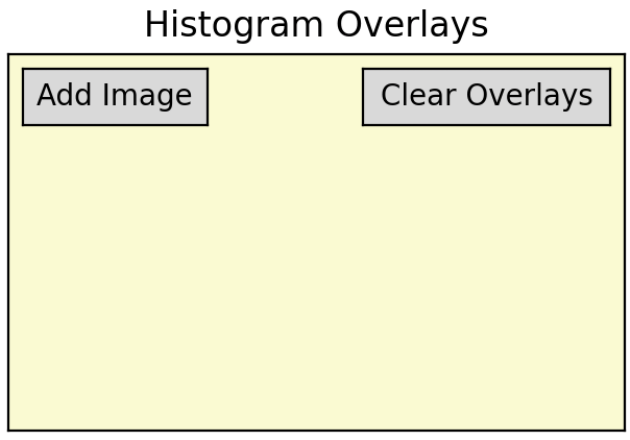} 
    \captionsetup{justification=raggedleft}
        \caption{The Histogram\\Overlays selection space,\\which contains two buttons,\\one which opens up a file\\dialogue to add images to\\the histogram and another\\which resets the workspace\\to its initial state with no\\overlays.}
    \label{fig:overlays-selection}
\end{figure}
\newpage
\begin{blockquote}
  program performance. If the user wishes to remove the overlays, one has to click on the “Clear Overlays” button in the fourth selection space, which will reset the histogram, the selection space, and the image displays to their original states.
\end{blockquote}
 \begin{figure}[H]
    \label{fig:overlays-singlge}
    
    \begin{center}
        \includegraphics[height=0.33\textheight]{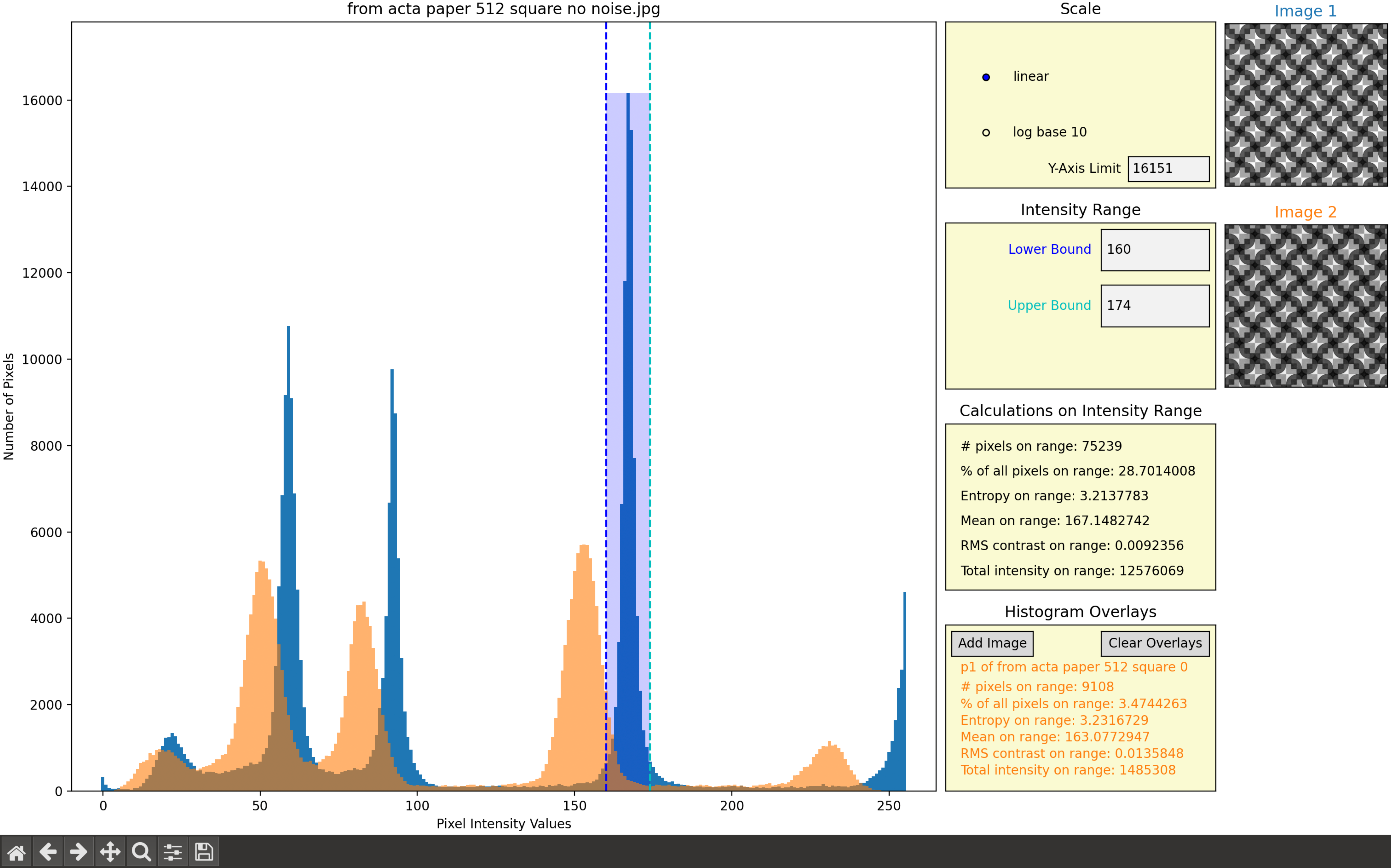}
    \end{center}
	\caption{The overlaid histogram, with the calculations in the "Histogram Overlays" selection space. The image corresponding to this histogram is displayed on the right under the title “Image 2,” with the font color matching the color displayed on the histogram and in the “Histogram Overlays” selection space. (This figure is displayed in color in the online version of this paper.)}
\end{figure}
\begin{figure}[H]
    \label{fig:overlays-multiple}
    \begin{center}
        \includegraphics[height=0.33\textheight]{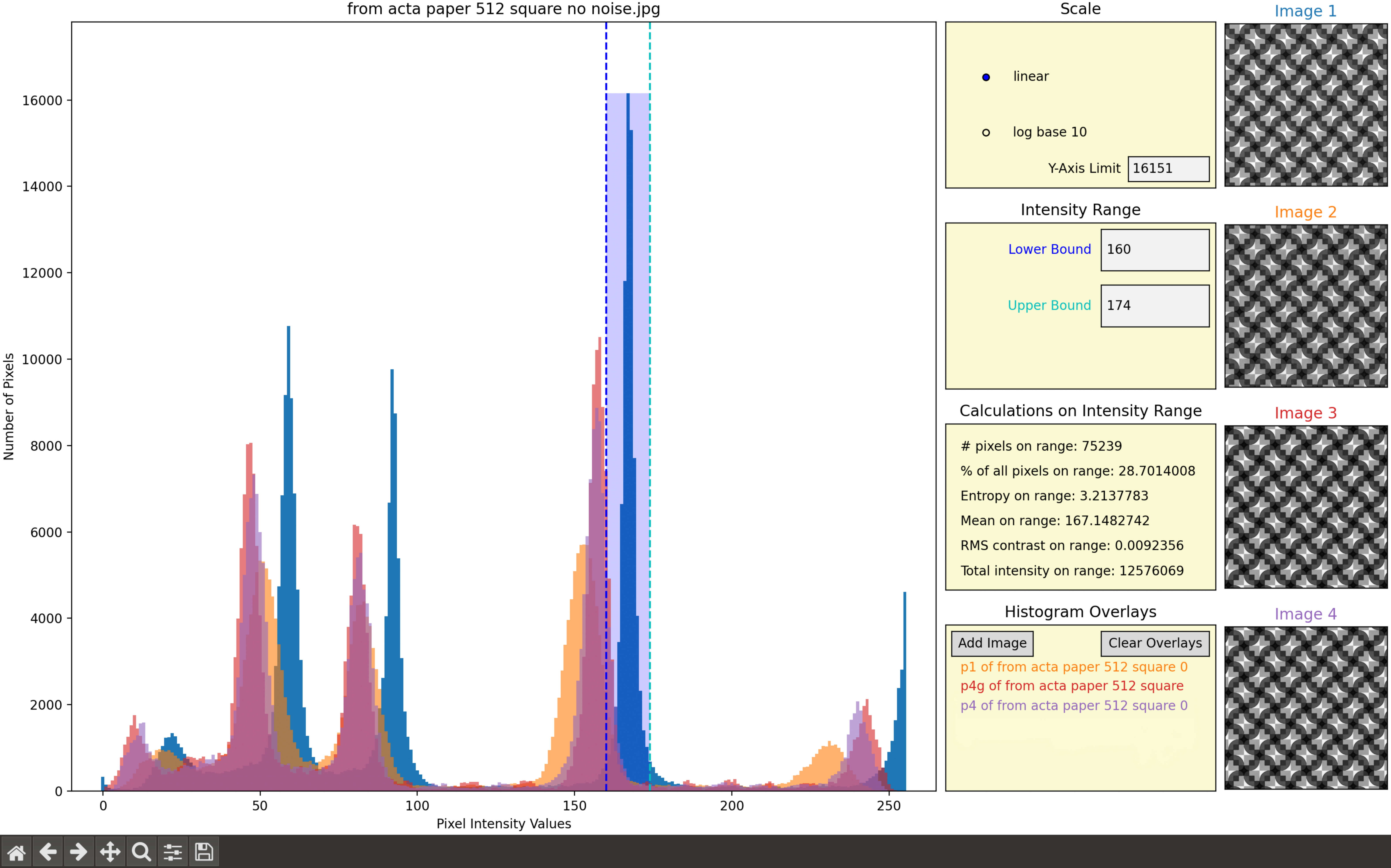}
    \end{center}
    \caption{Three overlaid histograms with the truncated file names in the "Histogram Overlays" selection space. Note that the images on the right correspond to the overlaid histograms, with the title color of each image matching the image that appears on the histogram and the filename that appears in the "Histogram Overlays" selection space. (This figure is displayed in color in the online version of this paper.}
\end{figure}
\newpage
\subsection{Navigating the Histogram}
The buttons in the bottom left corner of the Histropy workspace, Fig. 1, which appear underneath the histogram, allow the user to zoom in on specific parts of the histogram using the magnifying glass and move around the viewing window along the x- and y-axes with the axes button. When the user wishes to go back to a previous view, they can do so using the back arrow or they can move forward to the next view with the forward arrow. All viewing changes can be reset with the home button and the user can save the full plot image as a PNG with the save button.
\section{Concluding Remarks and Acknowledgment}
The functionality of the computer program Histropy has been briefly demonstrated in a few examples (based on versions of the crystal pattern in the appendix). An expanded description (handbook) of this program is freely available (together with the source code) in the supporting material for this paper. The authors hope that other researchers will find this paper interesting and eventually make good use of this program or parts of its source code in their own work.
The first author, the creator of the computer program, began this work and wrote initial drafts of this paper as a senior at Jesuit High School, and is currently a sophomore studying physics and computer science at the Georgia Institute of Technology. The second author is an applied crystallographer who has, in recent years, pioneered the application of a geometric form of information theory to crystallographic symmetry classifications of experimental data with two spatial dimensions. This project received support from a Faculty Development Grant of Portland State University. 
\section{References}

\begin{hangparas}{.25in}{1}
GIMP, 2024, https://www.gimp.org/.

GitHub, 2024, https://github.com/SMenon-14/Histropy. 

Hovmöller S., Oleynikov P. \& Zou X. (2011). \textit{Electron Crystallography. Electron Microscopy and Electron Diffraction.} Oxford University Press: Oxford Academic.

Ioannidis, Y. (2003). \textit{VLDB}. \textbf{29}, 19 ‒ 30. 

Kanatani, K. (1998). \textit{Intern. J. Computer Vision} \textbf{26}, 171 ‒ 189.

Moeck, P. (2022). \textit{Acta Cryst. A} \textbf{78}, 172 ‒ 199. 

Moeck, P. (2023). \textit{arXiv:2304.03915v3} (https://doi.org/10.48550/arXiv.2304.03915) and \textit{EasyChair Preprint № 10089}
      (https://easychair.org/publications/preprint/Cj97). 
      
Razlighi Q. R. \& Kehtarnavaz, K. (2009). \textit{Proc. of SPIE: Visual Communications and Image Processing} \textbf{7257},
      72571X-1 ‒ 72571X-10, DOI: 10.1117/12.814439.

Wood, C. (2026). \textit{arXiv:2602.07168v1}  (https://doi.org/10.48550/arXiv.2602.07168).
\end{hangparas}
\vspace{0.5cm}
\section{Appendix}
\renewcommand{\thefigure}{A\arabic{figure}}
\setcounter{figure}{0}
As an example of a subjective distinction between genuine symmetries and pseudosymmetries, the crystal pattern that forms the background in Figure A1a will likely be classified by most crystallographers at first sight as featuring the plane symmetry group \textit{p4gm}. However, upon closer visual inspection of the inset of Fig. A1a, which is a blow-up around site symmetry 2 at position 0, $\frac{1}{2}$ of the translation periodic unit cell, it becomes clear that this can only be a pseudosymmetry. Taking into account the intensity values of all of the pattern’s pixels, the information-theory-based crystallographic symmetry classification of this crystal pattern (Moeck, 2022) revealed that the genuine plane symmetry is indeed only \textit{p4}.  
In agreement with this objective classification, the “white bow-tie” feature of the inset does not genuinely feature point symmetry \textit{2mm} by visual inspection because the two diagonal mirror lines are broken to larger extents than the central two-fold rotation point. Moeck, 2022, and 2023 also give brief accounts of the creation of this pattern, from which it becomes clear that all mirror and glide lines can only be strong pseudosymmetries.
\newpage
\begin{figure}[htbp]
  \centering
  \begin{subfigure}[b]{0.45\textwidth}
    \centering
    \includegraphics[width=\textwidth]{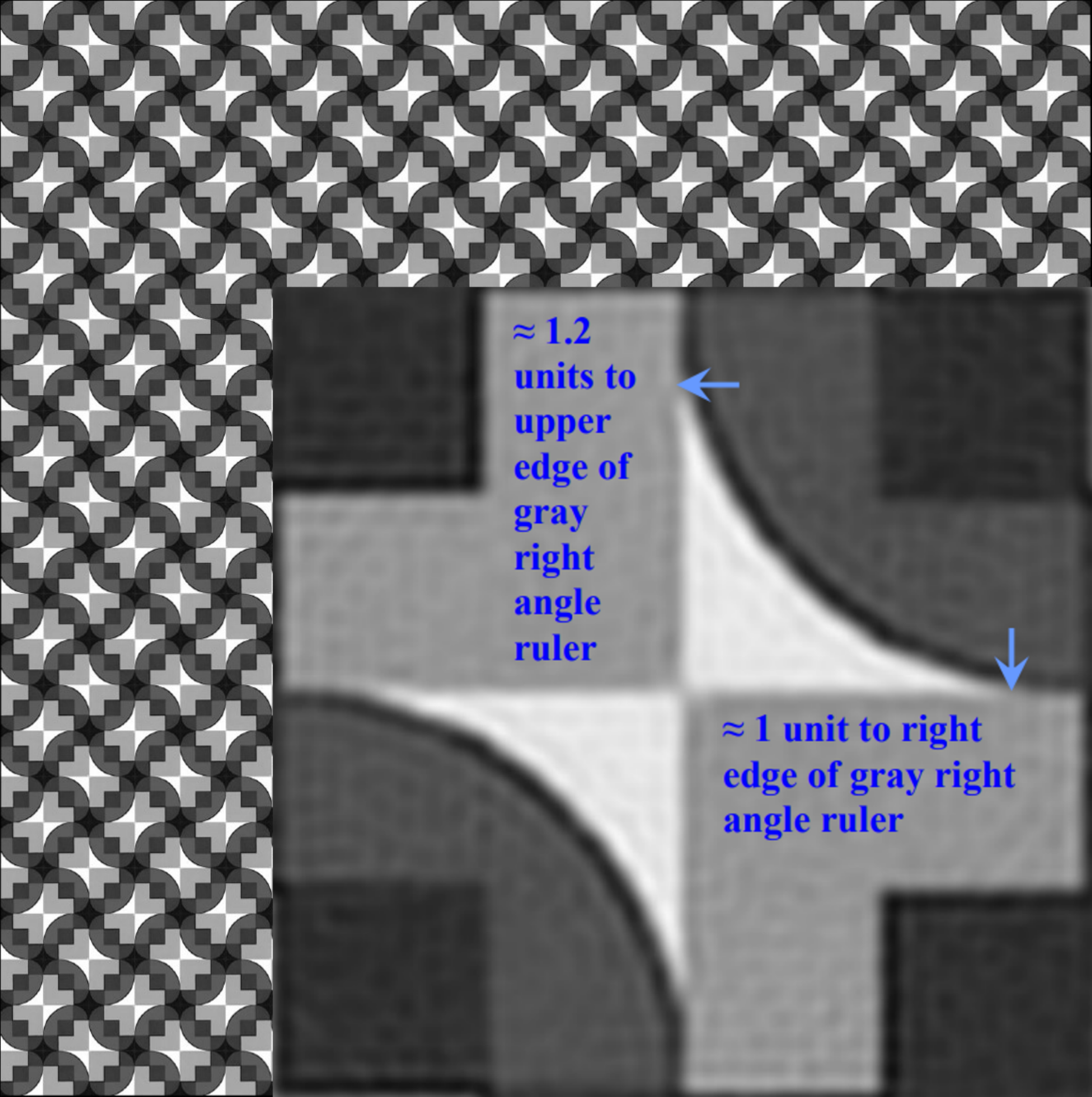}
    \caption{}
    \label{fig:acta-inset}
  \end{subfigure}
  \hfill
  \begin{subfigure}[b]{0.45\textwidth}
    \centering
    \includegraphics[width=\textwidth]{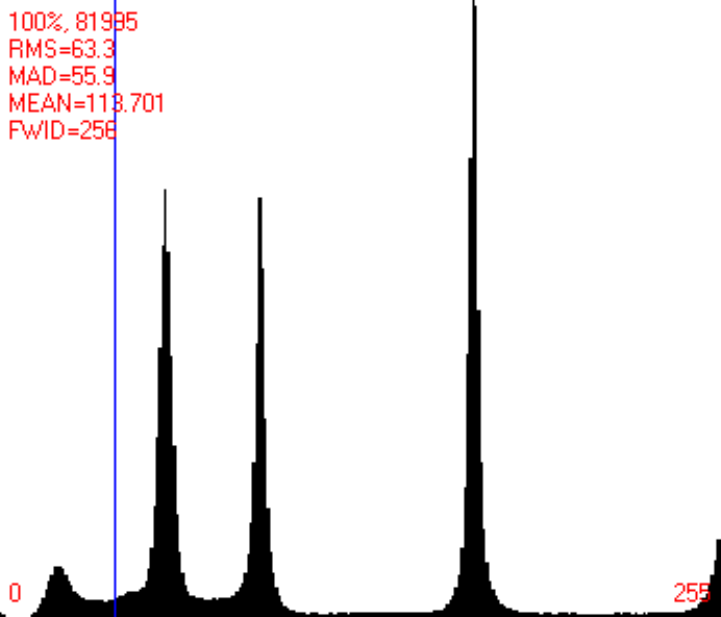}
    \caption{}
    \label{fig:CRISP-Hist}
  \end{subfigure}
  \caption{\textbf{(a)} This sub-figure displays a crystal pattern that will, by most people, be subjectively misclassified (at least at first sight) as featuring plane symmetry group \textit{p4gm}. The inset figure, which blows up a small part of the pattern, shows why this plane symmetry group must be dismissed as a pseudosymmetry. (Reproduced with permission from the paper and supporting materials of Moeck 2022.) \textbf{(b)} This sub-figure displays the histogram for the crystal pattern in A1a as obtained by the well-known electron crystallography program CRISP. (The figure is displayed in color in the online version of this paper.)}
  \label{fig:A1a}
\end{figure}
\noindent
Figure A2 illustrates the effect of the crystallographic processing of a noisy version of a 512 by 512 pixel cut-out of the crystal pattern in the background of Fig. A1a. Note that both the visual “sharpenings” and relative shifts of the histogram peaks of the two crystallographically processed version of the noisy image that were enforced to display the symmetries of the non-disjoint plane symmetry groups \textit{p2} and \textit{p4} are easily explained by the enhanced averagings over progressively smaller asymmetric units of the translation periodic unit cells. Note also that the histogram of the \textit{p4} (-enforced) version of the first (noisy) image is rather similar to those shown in Figs. 1 and A1b. This is a testament to both, the veracity of the image processing method and the correctness of the \textit{p4} plane symmetry classification of the noisy crystal pattern (Moeck, 2023). 
\begin{SCfigure}[0.95][h]
\label{fig:A2}
\includegraphics[width=0.53\textwidth]{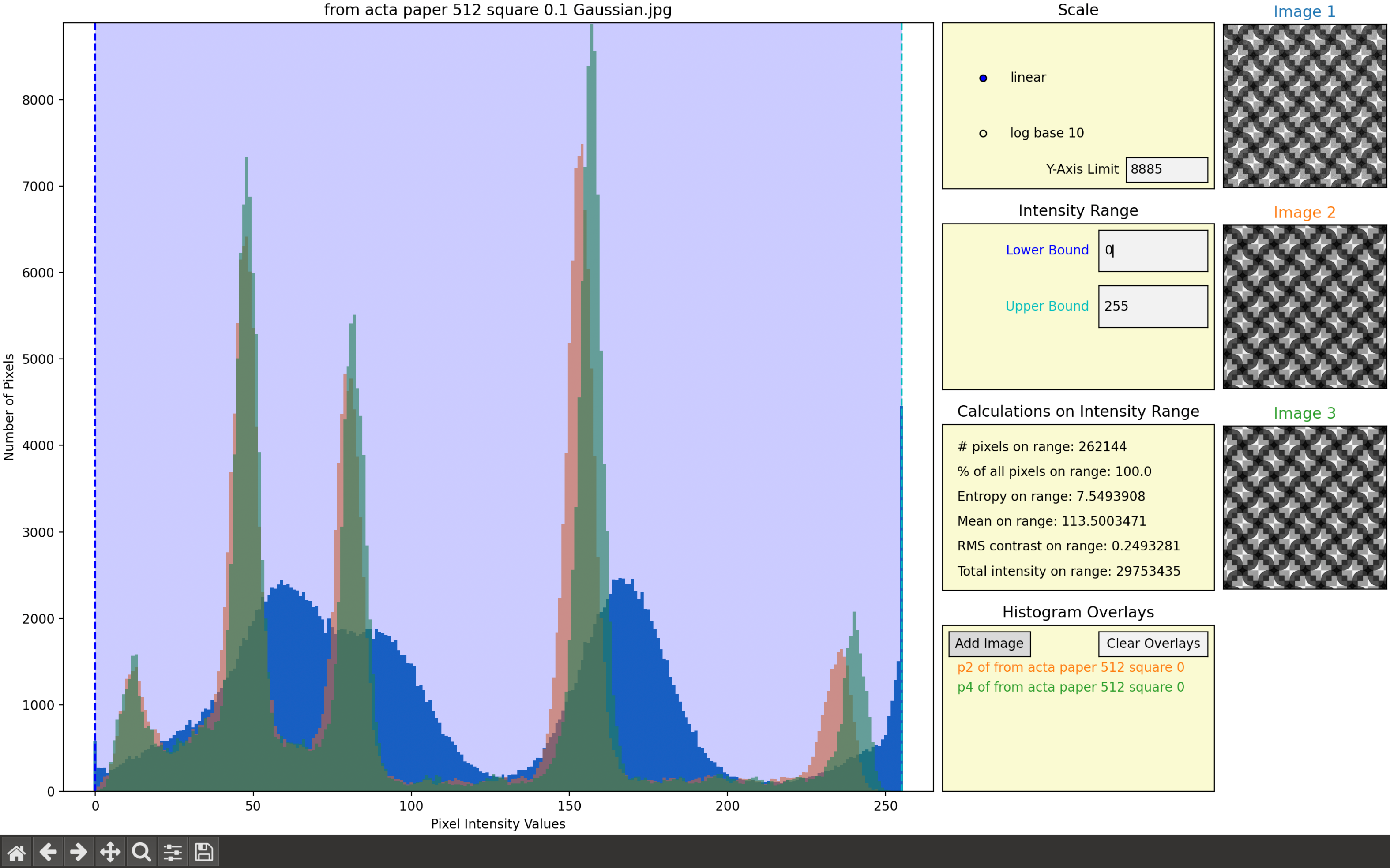}
\caption{Histograms of a noisy version of the crystal pattern in the background of Fig. A1a (displayed in blue) overlaid with the histograms of the \textit{p2} (displayed in orange) and \textit{p4} (displayed in green) image versions that were obtained by crystallographic image processing of that noisy image. (This figure is displayed in color in the online version of this paper.) The visual results, i.e. both peak sharpenings and shifts, in the histograms of the symmetry enforced versions of Image 1 are as expected.}
\end{SCfigure}

\end{document}